\documentclass[nofootinbib,aip,numerical,apl,reprint]{revtex4-1}

\usepackage{graphicx}
\usepackage{dcolumn}
\usepackage{bm} 
\usepackage{xcolor}
\usepackage[utf8]{inputenc}
\usepackage[T1]{fontenc}
\usepackage{mathptmx}
\usepackage{etoolbox}
\setcitestyle{numbers,square}
\makeatletter
\def\@email#1#2{%
 \endgroup
 \patchcmd{\titleblock@produce}
  {\frontmatter@RRAPformat}
  {\frontmatter@RRAPformat{\produce@RRAP{*#1\href{mailto:#2}{#2}}}\frontmatter@RRAPformat}
  {}{}
}%
\makeatother

\begin{document}
\preprint{AIP/123-QED}

\title{A Perspective on Electrical Generation of Spin Current for Magnetic Random Access Memories}

\author{Christopher Safranski}
\affiliation{IBM T. J. Watson Research Center, Yorktown Heights, New York 10598, USA}
\author{Jonathan Z. Sun}
\affiliation{IBM T. J. Watson Research Center, Yorktown Heights, New York 10598, USA}
\author{Andrew D. Kent}
\affiliation{Center for Quantum Phenomena, Department of Physics, New York University, New York, New York 10003, USA}%

\date{January 7, 2022}

\begin{abstract}
Spin currents are used to write information in magnetic random access memory (MRAM) devices by switching the magnetization direction of one of the ferromagnetic electrodes of a magnetic tunnel junction (MTJ) nanopillar. Different physical mechanisms of conversion of charge current to spin current can be used in 2-terminal and 3-terminal device geometries. In 2-terminal devices, charge-to-spin conversion occurs by spin filtering in the MTJ’s ferromagnetic electrodes and present day MRAM devices operate near the theoretically expected maximum charge-to-spin conversion efficiency. In 3-terminal devices, spin-orbit interactions in a channel material can also be used to generate large spin currents. In this perspective article, we discuss charge-to-spin conversion processes that can satisfy the requirements of MRAM technology. We emphasize the need to develop channel materials with larger charge-to-spin conversion efficiency---that can equal or exceed that produced by spin filtering---and spin currents with a spin polarization component perpendicular to the channel interface. This would enable high-performance devices based on sub-$20$ nm diameter perpendicularly magnetized MTJ nanopillars without need of a symmetry breaking field. We also discuss MRAM characteristics essential for CMOS integration. Finally, we identify critical research needs for charge-to-spin conversion measurements and metrics that can be used to optimize device channel materials and interface properties prior to full MTJ nanopillar device fabrication and characterization.
\end{abstract}
\graphicspath{{Figures/}}

\maketitle

\section{\label{Sec:Introduction}Introduction}
A widely studied spin transfer device is a magnetic tunnel junction (MTJ), a layered structure consisting of two ferromagnetic electrodes separated by a thin insulating tunnel barrier. An MTJ serves both to control the magnetic state, and to convert the magnetic state change into resistance differences.
For controlling the magnetic state, a bias voltage is applied to the junction, which leads to a spin-transfer torque on the magnetic layers associated with the flow of a spin-polarized current between the electrodes~\cite{Slonczewski1989,Slonczewski1996,Berger1996,Katine2000, Sun1999}. The magnetic state of the electrodes can be altered if a sufficient amount of spin-polarized current is supplied. An MTJ's magnetic state can be determined by electronic means, as the MTJ's tunnel resistance depends on the relative magnetization orientation of the two tunnel electrodes --- due to the tunnel magnetoresistance effect (TMR)~\cite{Maekawa1982,Moodera1995,Butler2001, Yuasa2004,Parkin2004}. Thus, at a small bias voltage, the MTJ is used to read out the magnetic state, while at higher bias voltages, its magnetic states can be altered. Such an MTJ device forms the basis for spin transfer torque magnetic random access memories (STT-MRAM), now being widely developed worldwide by the semiconductor industry~\cite{Kent2015,Slaughter2012,Rizzo2013,Thomas2015,Slaughter2017}. Improvements in efficiency and new ideas for generating spin torques can thus have a huge impact on this nascent semiconductor memory technology.

The fundamental mechanism for switching a nanomagnetic MTJ electrode in this case is the flow of spin angular momentum carried by tunnel electrons. The basic physics is tunnel electron's spin dephasing, and its related conservation of angular momentum: spin angular momentum in the itinerant electron system (the electric current) can be exchanged with that of the magnetization of the electrodes~\cite{Slonczewski1989,Slonczewski1996,Berger1996,Stiles2002,Slonczewski2005}. For an MTJ, spin-polarized current is derived from the fact that ferromagnetic metals have a spin-dependent electronic structure with different densities of states at the Fermi level of majority (spins aligned with the net angular momentum in the magnetic electrode) and minority spins~\cite{Slonczewski1989}~\footnote{For a review of earlier research on spin-polarized tunneling see~\cite{Meservey1993}.}. 

Spin-polarized tunneling by an MTJ is only one of many approaches to convert a charge current into a spin current. More recently, it has been shown that large spin-polarized currents and spin accumulations can also be generated by current flow in non-magnetic layers with strong spin-orbit coupling~\cite{Miron2010,Liu2012} through, for example, Rashba and spin-Hall effects~\cite{Hirsh1999,Zhang2000}. This provides an alternative approach for charge-to-spin current conversion\footnote{We will often simply use the terminology ``conversion efficiency'' to refer to charge-to-spin current conversion efficiency, unless we indicate otherwise.}, with potentially higher conversion efficiency than spin-polarized tunneling. Materials with strong spin-orbit coupling combined with MTJ-based read out can open new applications for memory technology and beyond. For high-density commercial memory applications, the efficiency of the charge-to-spin current conversion in either case determines the device efficiency, and thus its technological viability. 

This perspective article highlights current approaches and challenges to electrical charge-to-spin conversion and, in particular, controlling the spin polarization direction using spin-orbit coupling in magnetic materials. 
This conversion is benchmarked to the conversion efficiency in MTJs and the industry requirements for advanced technology nodes.

\section{Device Physics}
\begin{figure}[t]
  \centering
   \includegraphics[width=1\columnwidth]{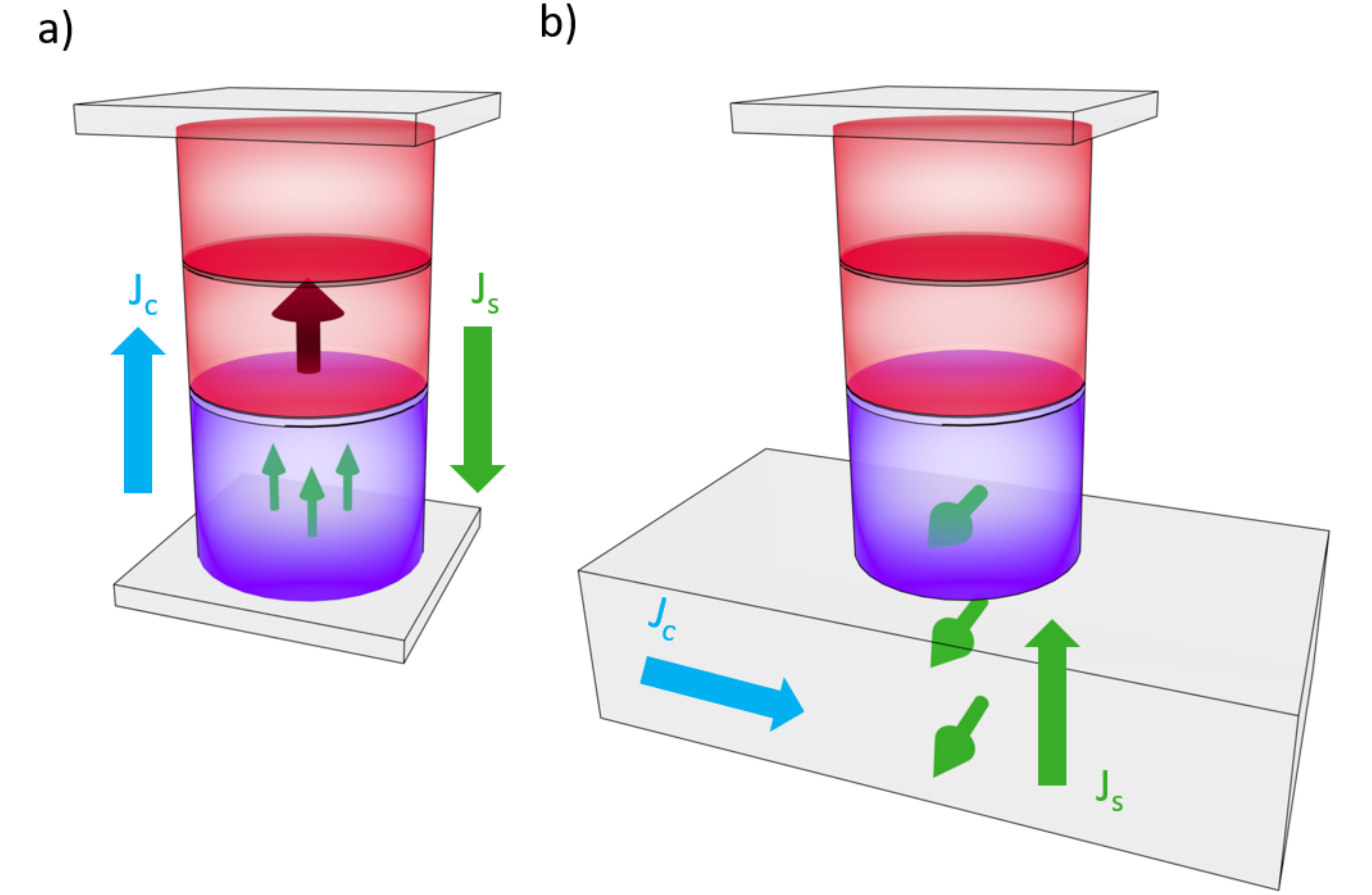}
    \caption{\textbf{Charge-to-spin conversion and magnetic device geometries.} The free layer is indicated in purple and the reference layer in red. a) Spin-polarized tunnel current in a 2-terminal pMTJ nanopillars; the spin current flow (green arrow) is collinear with he charge current (blue arrow), perpendicular to the free layer plane. The arrow in the reference layer indicates its magnetization direction. b) Spin-orbit-interaction spin current generation in a 3-terminal magnetic device.  Here the charge current flows in a channel (indicated in gray) and the spin current flow into the free layer is perpendicular to the plane of the free layer.}
\label{Fig:Devices}
\end{figure}

For applications, two frequently examined methods of sourcing and delivering spin current are those of a spin-polarized tunnel current approach, and those based on spin-orbit interactions. Figure~\ref{Fig:Devices} shows schematics for both a 2-terminal (Fig.~\ref{Fig:Devices}(a)) and 3-terminal (Fig.~\ref{Fig:Devices}(b)) spin transfer device. 

In a 2-terminal (and often perpendicular magnetized MTJ -- or pMTJ\footnote{The importance of pMTJ is derived from its higher achievable magnetic anisotropy energy density, which enables higher areal-density integration with circuitry for memory applications, because the anisotropy energy here is directly related to a memory's data-retention lifetime. In practice the total anisotropy of a nanomagnetic ``bit'' needs to be of the order of 60$k_B T$ or higher for 10 year data retention at the operating temperature $T$. Further, a uniaxial perpendicular anisotropy also minimizes the spin-current switching threshold for a given anisotropy barrier, reducing required switching charge current, which is critically important for high-density memory technologies.}) nanopillar, a current flows through a thin insulating tunnel barrier perpendicular to the plane of the layers. The current is spin polarized and the spin and charge current flow parallel to the same axis in opposite directions. The spin current exerts spin torques on the layers and the layers are designed such that one layer (the ``free layer'') magnetization switches it magnetization orientation with respect to that of a reference magnetic layer. 

In a spin-orbit interaction based 3-terminal device, the charge current flows in the plane of layer in a channel (horizontally in Fig.~\ref{Fig:Devices}(b)) and spin-orbit coupling can generate a spin current flow perpendicular to the current propagation direction (hence the term ``spin-Hall effect''); this spin current can switch the magnetization direction of a free magnetic layer that is connected to the channel with a good spin-conduction contact. In both situations, a spin current is characterized by two directional vectors (i.e. it is a rank-2 tensor), one its spin-polarization direction, the other the spin-current propagation direction. 

\subsection{Charge-to-spin conversion}
In both tunnel spin filtering and in spin-orbit based spin-current generation, a key metric is the charge-to-spin current conversion efficiency. This is given by the dimensionless ratio known as the spin-Hall angle $\theta_\mathrm{SH}=J_s/J_\mathrm{c}$, where $J_\mathrm{c}$ is the charge current density, and $\hbar J_\mathrm{s}/(2e)$ is the spin current (i.e. $J_\mathrm{s}$ is the spin-current written in units of charge current density by replacing an electron's spin $\hbar/2$ by its charge, $e$, where $\hbar$ is Planck's constant and $e$ is the magnitude of the electron's charge. The two different device geometries shown in Fig.~\ref{Fig:Devices} differ with respect to the flow of angular momentum. In the case of the 2-terminal pMTJ, the charge and spin currents flow through the same surface area (perpendicular to the layer planes in the nanopillar) and the ratio of the spin current to the charge current, $I_s/I_c$, is equal to $J_s/J_\mathrm{c}$. By contrast, in an SOT-based device, the spin current flows primarily perpendicular to the charge current and parallel to the normal direction of the thin film interfaces. Thus the total spin- and charge-current accounting relates to different geometrical areas. As a result, the ratio of spin current to charge current is $I_s/I_c=
\theta_\mathrm{SH} (\ell/t)$, where $\ell$ is the lateral size of the nanopillar and $t$ is the thickness of the layer generating the spin current. The geometrical factor $\ell/t$, is usually greater than $1$ and can thus serve to augment the spin current. It is also important to note that there are several factors that can reduce the efficiency of the angular momentum transfer, such as loss, per reflection or absorption, of spin angular momentum at interfaces, characterized by an spin transparency factor and a so-called interface spin-memory-loss factor~\cite{Zhu2021b,Bass2016}. Furthermore, as the spin density is not conserved in conductors, it is characterized by both a diffusion constant and a spin lifetime.

\subsection{Switching current}

The conservation of angular momentum provides a useful way of thinking of spin-transfer induced magnetization dynamics and can be used to make an estimate of the spin current needed to switch the free layer magnetization and write information. In order of magnitude, the minimal total number of spins incident on the free layer---the spin current times the pulse time---for switching with short duration ($\sim$ns) pulses is related to the number of elemental spins or Bohr magnetons associated with the free-layer magnetic moment. Thus the larger the charge-to-spin conversion ratio, the smaller the charge current times the pulse duration (i.e. the total number of charges) needed to write information. The spin polarization direction relative to the free layer magnetic easy axis is also important. When the two are collinear, the spin torque can fully oppose the damping\footnote{References~\cite{Sun1999,Sun2000,Mancoff2003,Garzon2008} consider the case in when the spin polarization and the free layer's magnetization are not collinear.}, which is known as ``antidamping'' switching. Here for one flow direction the spin current can amplify fluctuations and deviations of the magnetization from a collinear alignment, leading to switching. When spin is polarized orthogonal to the magnetic easy axis, the situation changes and, while the switching can be faster, larger currents are generally required, as we discuss further below in Sec.~\ref{Sec:OSTS}.

\subsubsection{Switching instability threshold}
\label{Sec:SIT}
\begin{figure}[t]
  \centering
   \includegraphics[width=1\columnwidth]{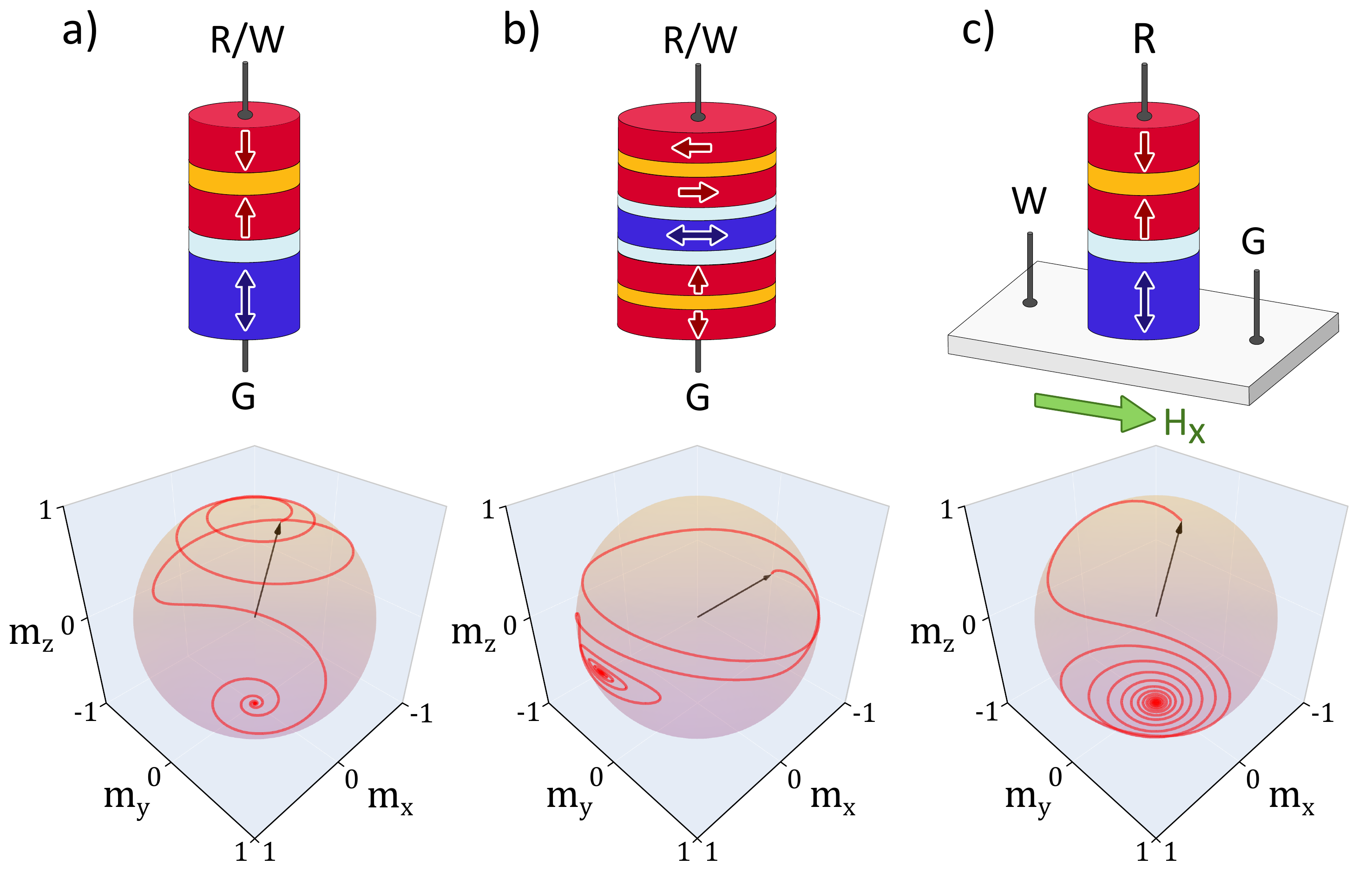}
    \caption{\textbf{Antidamping and orthogonal spin transfer switching.} a) Antidamping switching for a free layer with perpendicular magnetic anisotropy. In this case, for one current polarity, the STT opposes the damping leading to magnetization reversal. b) Orthogonal spin torque for an in-plane magnetized free layer. The reversal is by magnetization precession about the hard axis with the spin-polarization angle, pulse polarity, amplitude and duration determining the final magnetization state. c) Orthogonal spin torque for a free layer with perpendicular magnetic anisotropy. A symmetry breaking interaction---such as an applied in-plane magnetic field---is required for deterministic switching. The read, write and ground device terminals are indicated by ``R'',''W'' and ''G'', respectively.}
\label{Fig:Orthogonal}
\end{figure}

\paragraph{Perpendicularly magnetized layers.}
\label{Sec:PMA}
The switching of a nanomagnet can be analyzed using magnetodynamics described by the Landau-Lifzhitz-Gilbert-Slonczewski equation with the damping-like spin-transfer torque included~\cite{Slonczewski1996,Sun2000,Kent2021}. We start by examining the switching of free layer with strong perpendicular magnetic anisotropy by the antidamping switching scenario (as illustrated in Fig.~\ref{Fig:Orthogonal}(a)). In this case, the spins are polarized perpendicular to the film plane and thus collinear with the magnetic easy axis of the free layer. In this geometry, and in the macrospin limit---in which magnetization of the free layer is assumed to be always spatially uniform---the threshold for antidamping switching at zero temperature is directly proportional to energy barrier to thermally induced reversal, $E_b$~\cite{Slonczewski1996,Sun1999,Sun2000}, the magnetic anisotropy barrier that separates the magnetization up and down states:
\begin{equation}
V_\mathrm{c0} = \left ( \frac{4e}{\hbar} \right ) \left (\frac{\alpha}{\eta G_P } \right ) E_b,
\label{Eq:Vc}\end{equation}
where $\alpha$ is the damping of the free layer and $G_p$ is the MTJ conductance in the parallel state in the low-bias limit. $\eta$ is the effective charge-to-spin current conversion coefficient {\it when the MTJ is in the parallel state}, $\eta=\sqrt{m_r\left( m_r+2\right)}/\left[2\left(m_r+1\right)\right]$, where $m_r$ is the junction's magnetoresistance, related to the resistance of the junction in the antiparallel $R_\mathrm{AP}$ and parallel $R_\mathrm{P}$ magnetization states as $m_r=(R_\mathrm{AP}-R_\mathrm{P})/R_\mathrm{P}$. The expression for $\eta$ assumes a symmetric junction, with the same electrode materials and identical interfaces on each side of the tunnel barrier (see Ref.~\cite{Slonczewski2005,Slonczewski2007}). 

Eq.~\ref{Eq:Vc} shows that the switching threshold is directly proportional to the free layer's damping and to leading order is inversely related to the junction's magnetoresistance. Of note, the threshold is relatively insensitive to the magnetoresistance for $m_\mathrm{r}>1$, corresponding to 100\% TMR, i.e. further increases in the magnetoresistance do not greatly reduce the threshold voltage.

An important consequence of Eq.~\ref{Eq:Vc} is that for MTJs with large $m_r$, the charge-to-spin {\it current conversion ratio} is dependent on the relative orientation of the magnetization's of the electrodes, while the {\it voltage to spin-current relationship} is not. This is due to the different sub-channel summations for spin- and charge-current. See Refs.\cite{Slonczewski2005,Slonczewski2007,2016192,2021050} for in-depth discussion of this point and further references. As a consequence, the switching threshold voltage $V_\mathrm{c0}$ for antiparallel-to-parallel (AP-P) switching and for P-AP (where the initial moment alignments are  $180^\circ$ different) are identical, whereas the P-AP threshold charge current is higher than that of the AP-P switching instability threshold. For convenience in circuit applications, it is common to refer the P-AP instability charge current threshold $I_\mathrm{c0}=V_\mathrm{c0} G_p $ as the ``switching current'' metric of an MTJ, even though the AP-P instability threshold current would be lower. 

In terms of conversion efficiency, for large TMR,  $m_r \rightarrow +\infty$, the related charge-to-spin conversion ratio is $\eta=1/2$~\cite{2016192,2021050}. It is $1/2$ because the resulting spin torque is equally shared by the two electrodes’ interfaces.  Therefore, $\eta > 0.4$ as a reference for what has been achieved in 2-terminal spin-transfer MTJ devices, is not far from the ideal limit of $1/2$. It also set a benchmark---{\em any novel device physics implementations must demonstrate an advantage against this charge-to-spin conversion ratio.}

\paragraph{In-plane magnetized layers.}
\label{Sec:IMA}
Another important case is for a MTJ with in-plane magnetized electrodes, again in the presence of an antidamping spin transfer torque. In this case the free layer has a biaxial magnetic anisotropy, an easy axis in the film plane and a hard axis perpendicular to the film plane; the magnetic states are typically stabilized by magnetic shape anisotropy, by having a free layer element with an elliptical shape. The shape anisotropy can be characterized by an in-plane magnetic anisotropy field $H_k$, giving an energy barrier $E_b=\mu_0M_sH_k \mathcal{V}/2$, where $\mathcal{V}$ is the volume of the free layer. However, the switching voltage and current in a macrospin model are increased relative to that in Eq.~\ref{Eq:Vc} by a factor of $(1+D)>2$, where $D$ is the ratio of the easy-plane to easy-axis anisotropy of the free layer $D=M_\mathrm{eff}/H_k>1$, where $M_\mathrm{eff}$ characterizes the easy-plane magnetic anisotropy\footnote{For thin free layer, i.e. a layer with lateral dimension much larger than its thickness, $M_\mathrm{eff}$ is the ferromagnet's magnetization density minus the perpendicular magnetic anisotropy field $H_p$, i.e. $M_\mathrm{eff}=M_s-H_p$. See, for example, Ref.~\cite{Chaves2015} for further details. }~~\cite{Sun2000,Pinna2013,Pinna2014,Sun2016}. While the threshold voltage and current increase relative to the case of perpendicularly magnetized layers, the impact on the overall switching-current requirement for a given write speed and write-error-rate, however, could be less significant. This is because to obtain very low switching errors, a fast write requires a spin current that is much larger than the instability threshold, and the switching dynamics can exhibit non-macrospin behavior~\cite{Sun2013,Sun2016,Bouquin2018,Volvach2020,Mohammadi2021} (e.g. non-uniform magnetization reversal and micromagnetic instabilities) that can modify the difference an increase in instability threshold makes to the overall memory cell optimization.

\subsubsection{Switching current at finite temperature}
For most memory applications, devices need to operate near room temperature. This places several vital requirements for STT-switched magnetic memory. These factors determine the often challenging trade off between the need to minimize switching (aka ``write'') current on one hand, and three main technology performance metrics on the other:  write errors, switching speed, and data retention lifetime.

First, a nanomagnet based memory always has a finite thermal activation energy $E_b$ between ``0'' and ``1'' states and thus has a small but finite probability to thermally activate over the energy barrier stochastically, causing a random memory error. The error rate of this process determines the mean data retention time of the memory. Typical non-volatile memory technology requires a reliable data retention lifetime on an entire chip to better than 10 years. For typical nanomagnet materials used in an MTJ memory arrays, this corresponds to requiring $E_b \gtrsim 60 k_B T$, with $T$ the operating environment's temperature, for 10 year error free operation of a Gb memory chip\footnote{This is dictated by the Arrhenius thermal activation rate $\tau \sim \left(1/\gamma H_k \alpha\right) e^{E_b/k_B T}$, with $\gamma=2\mu_B/\hbar$ the gyromagnetic ratio, $H_k$ the uniaxial anisotropy field, and $\alpha$ the Gilbert damping. We note that $E_b \gtrsim 40 k_B T$ is sufficient for a single bit to be stable for more than 10 years. However, in a large memory array larger bit stability is required to maintain stable memory states of the entire array. To account for distributions of magnetic device characteristics within the array, values even much larger than $60 k_B T$ may be required.}. Note this specific error mechanism of thermal-activation induced reversal is a constant error rate, i.e. a constant error probability per unit time. It is different from the data-retention time of capacitance-storage based memory, such as dynamic random-access memory (DRAM). The constant error rate of magnetic memory can be addressed by error correction methods, but cannot be refreshed away by rewriting~\cite{2021159}.

The thermal activation barrier height $E_b$ sets an energy scale that governs the minimum ``write'' spin-current level. As indicated by Eq.~\ref{Eq:Vc}, the threshold for STT switching in {\em spin current} is, in the macrospin limit, $I_\mathrm{s0} = \left(4e/\hbar\right) \alpha E_b$. 

Secondly, thermal fluctuation of the nanomagnet also means a probabilistic distribution of the initial moment position when the switching pulse is applied. A Boltzman-distribution estimate relates the mean initial azimuthal angle $\theta$ (with respect to the magnetic easy-axis, such as film normal in a pMTJ) to $E_b$ as $\left< \theta^2 \right> \sim\left({k_B T}\right)/{E_b}$~\cite{2016084}. For a given spin-current $I_s > I_\mathrm{s0}$, a nanomagnet switches faster for larger initial $\theta$~\cite{Sun2000,Liu2014}. Therefore, a distributed initial angle $\theta$ means the magnetization switching time becomes a distributed statistical quantity. Consequently, at a given write pulse duration, $\tau_w$, an MTJ would switch probabilistically, with the switching probability increasing with increasing spin-torque drive amplitude $I_s$. Thus an increase in switching current is a requirement for faster and more reliable switching. For example, for some 2-terminal STT-pMTJ devices switching with an write error rate (WER) of $10^{-6}$ in 3\,ns requires a switching current that is 50\% larger than the threshold\footnote{The write error rate is defined as the probability of failing to write per write operation.}~~\cite{2021005}. Much of the 2-terminal STT-MTJ based MRAM development involves the balance of data retention related to $E_b$, and the switching current for fast, reliable switching, while fitting into the targeting technology node's transistor size-related current and impedance requirements. Since the relevant switching current here is a spin current, this drives the need for ever greater charge-to spin-current conversion efficiency.  For a 2-terminal STT-MTJ based memory cell, a further need for charge current reduction is related to device endurance to repeated write operation. This arises because one needs to stay well below the current density limit and voltage limit for tunnel barrier breakdown. This limit could be mitigated by separating charge and spin-current path in three terminal memory devices, as discussed later in this article.

Lastly, thermal fluctuations can also be magnified by MTJ voltage bias below the instability threshold. In this region, a spin current due to finite MTJ bias voltage would increase the probability of thermal activation over the barrier, resulting in memory state error. This becomes more serious as one approaches the threshold instability current or voltage~\cite{2016084,Sun2016}. In practice, this is one factor setting the applied bias for reading the state of the MTJ to prevent what are known as read-disturbs during device operation.

\subsubsection{Orthogonal spin transfer switching}
\label{Sec:OSTS}

Spin transfer switching when the spin polarization and accumulation are orthogonal to the magnetization has characteristics that are distinct from antidamping spin torque switching---the case discussed in the above sections when the spin polarization and accumulation are collinear with the magnetization. Orthogonal spin-transfer (OST) switching depends on many details of the magnetic layer configurations and the symmetries of the magnetic interactions. There are several cases that have been explored in some depth both in modeling and experiment: 1) an in-plane magnetized free layer, a layer with an easy axis in the film plane and a hard axis perpendicular to the film plane and 2) a perpendicularly magnetized free layer, a free layer with a predominantly easy axis type magnetic anisotropy.

Figure~\ref{Fig:Orthogonal}(b-c) shows two cases of OST switching. For an in-plane magnetized free layer (Fig.~\ref{Fig:Orthogonal}(b)), a spin current greater than a threshold leads to magnetization precession about the hard axis, with the threshold current set by the in-plane easy axis magnetic anisotropy. The threshold current is much larger than the case of antidamping switching (by a factor of order of $1/\alpha$), because the spin-current threshold is proportional to the magnetic anisotropy rather than the damping times a magnetic anisotropy (e.g., in Eq.~\ref{Eq:Vc}). An advantage of this OST
configuration is that switching can be faster. A disadvantage is the larger spin currents are needed. In addition, the spin-polarization direction, pulse amplitude and duration are critical variables in determining the final magnetic state~\cite{Kent2004,Rowlands2019}. We note that when the spin polarization has a component along the easy axis (i.e. it is not completely perpendicular to the magnetization) the threshold current decreases and the switching dynamics depends on the angle of the spin polarization, the pulse polarity and, again, the pulse shape.

The second SOT case (Fig.~\ref{Fig:Orthogonal}(c)) of an in-plane spin polarization and perpendicularly magnetized free layer represents switching of a perpendicularly magnetized free layer by spin-orbit torques generated by heavy metal layers (Fig.~\ref{Fig:Orthogonal}(c)). In this case, a symmetry breaking interaction is needed for deterministic switching (i.e. switching in which the current polarity determines the final magnetic state of the free layer). This can be provided by an applied in-plane magnetic field, an internal field or a second magnetic layer. Again, in this case the threshold current for switching is generally larger than that for antidamping spin torque switching by a factor of approximately $1/\alpha$. 

There are several factors that can lower the threshold current and make the switching deterministic. A device geometry that combines antidamping and OST can lower the threshold current and provide a deterministic switch. Modeling has indicated that spin currents that produce effective field interactions on the free layer can lead to deterministic switching~\cite{Taniguchi2015a}. Finally, a spin-polarization that is canted relative to the film plane provides an antidamping component to the torque that can make the switching deterministic.

\subsection{Spin current sources: advantages and disadvantages}
\label{Sec:SCS}

\begin{widetext}

\begin{table}
\caption{Comparison of key features of existing and emerging magnetic memories.  A 2-terminal device requires 1 access transistor, indicated as 1T. A 3-terminal device requires 2 transistors (2T), one for writing and one for reading. The terminals for these transistors are labeled "W" for the write and   "R"  for the read transistor connections which pass current to a ground connection "G."} 
\label{Table:MM}   
\includegraphics[width=0.9\textwidth]{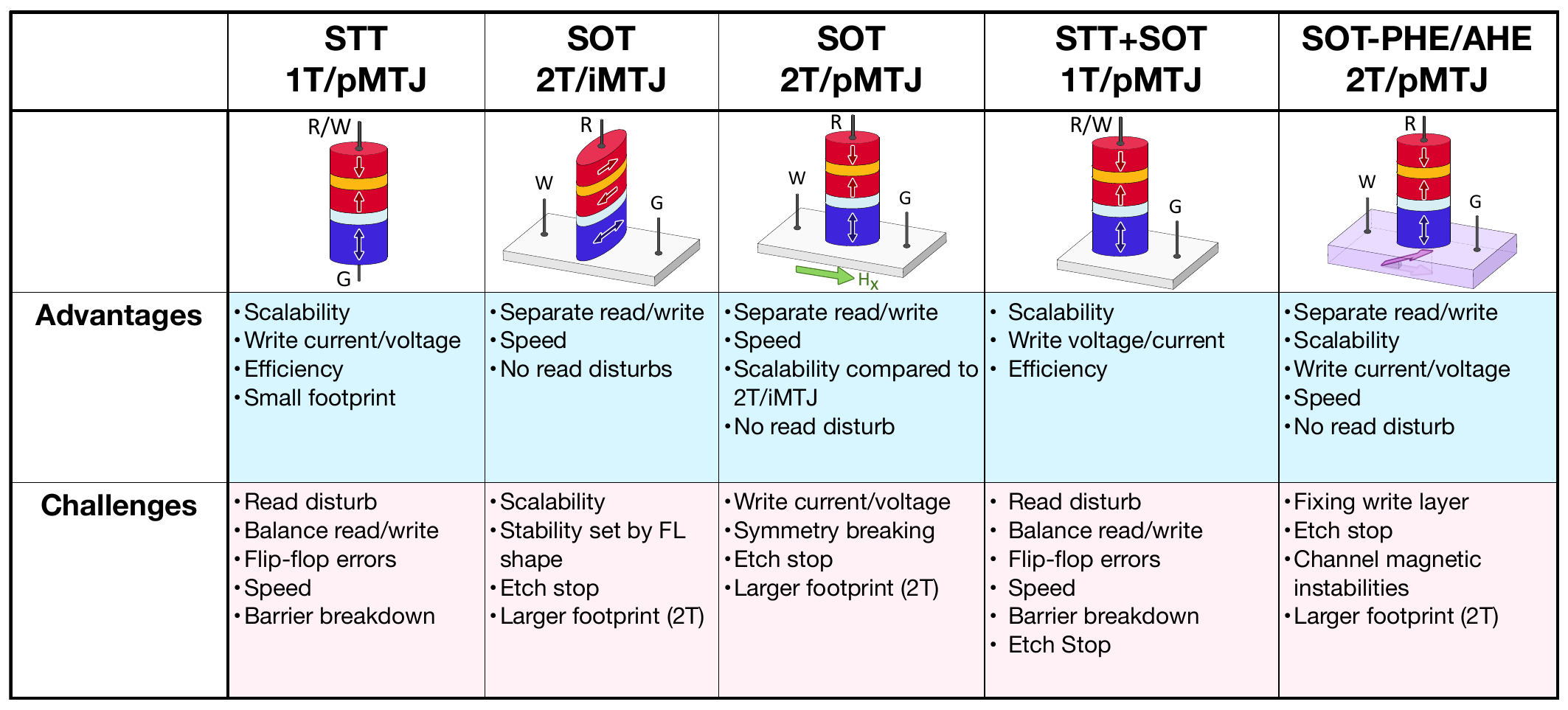}
\end{table}
\end{widetext}

The  possible MRAM write mechanisms utilizing different spin current sources  come with different advantages and challenges for making them viable for commercial applications. These are summarized in Table~\ref{Table:MM}. Currently available  MRAM technology is based on the so-called 1-transistor, 1-pMTJ memory cell (or a 1T/pMTJ cell), using a 2-terminal pMTJ magnetic element with the same current path for reading and writing, to maximize bit density. For pMTJ, switching currents have been optimized for matching the transistor current, typically below $10^7$A/cm$^2$ for reliable (WER $<10^{-6}$) and sub-5ns writing~\cite{Hu2019}. This is lower than the best SOT write current demonstrated so far. As discussed in Sec.~\ref{Sec:SIT}, there is a fundamental limit to increasing charge-to-spin conversion efficiency of the 2-terminal MTJ because this approach utilizes spin filtering. The simple 1T/pMTJ design also requires co-optimization of read and write circuits, which requires some compromise to both functions.

The are also technical issues with using the same current path for read and write operations. In order to read the magnetic state of the MTJ, a current must be passed through the device which can possibly switch the free layer causing a read disturb error. Further, in the write process, passing large currents across the tunnel barrier leads to stress, and a corresponding barrier integrity related reliability consideration that needs to be properly managed. Additionally, reference layer instabilities related to ``flip-flop'' or ``back-hopping'' errors can appear at high currents~\cite{Kim2008,Min2009,Kim2016,Abert2018,Safranski2019,Devolder2020}, possibly limiting the lowest write error rate that can be achieved.  These  engineering challenges  constrain the application space for the 1T/pMTJ cell. Thus the need to explore other cell designs, especially for higher-speed spaces, such as last-level cache replacement.

Moving to a SOT-based 3-terminal 2-transistor (2T) design is one possible way to mitigate the constrains facing the 1T/pMTJ cell. The switching current is now sourced from a separate write channel below the tunnel junction. This allows the read/write paths to be separately tuned, and eliminates the problems with read disturb, high-bias barrier breakdown, and flip-flopping. However, one  problem with this approach that needs to be addressed for commercial applications is the patterning process involved in defining the MTJ and write channel. The material thickness of the write channel is often on the order of a few nanometers and the channel/free layer interface needs to be extremely well controlled, requiring the interface to be formed in situ with the MTJ. Thus the channel generally needs to be below the free layer (as indicated in the illustrations in Table~\ref{Table:MM}). This means that successful MTJ nanopillar definition requires a tight control of the MTJ etch conditions across an entire wafer, to stop the etch just below the free layer, and not significantly etch away the channel material.

So far for SOT based structures, the largest charge-to-spin conversion efficiency has been achieved using spin-orbit effects that produce an in-plane spin polarization. As discussed in Sec.~\ref{Sec:SIT}, magnetic switching with minimum spin-current requirement is achieved with antidamping action, which means the spin-polarization needs to be collinear with the magnetic easy-axis. Since the spin polarization from spin-orbit action in an isotropic channel material is in the substrate plane and perpendicular to the current propagation direction ($\mathbf{n}_s = \mathbf{n_\mathrm{c}} \times \mathbf{n_f}$, where $\mathbf{n}_s$ is the spin-polarization direction, $\mathbf{n}_\mathrm{f}$ the film normal, and $\mathbf{n}_\mathrm{c}$ the charge current propagation direction), a collinear alignment favors an in-plane easy-axis direction for the memory bit. This device geometry is illustrated in Table~\ref{Table:MM} under the heading SOT 2T/iMTJ. In this case, the free layer must be relatively thin to have a low antidamping switching threshold current (which is proportional to the element volume). This, with other practical considerations (including the ability to control the lateral shape of elements across a wafer), indicate that an in-plane magnetized MTJ memory bit will not scale well below $\sim 100$ nm in lateral size, limiting its use in high-density commercial applications\footnote{Specialty applications remain viable for such cell designs, as was demonstrated by Everspin's successful first generation STT-MRAM product offering~\cite{Slaughter2012}.}.

To decrease the device footprint, it is desirable to take the 2T design together with a perpendicularly magnetized MTJ (a 2T/pMTJ cell), shown in Table~\ref{Table:MM} under the heading SOT 2T/pMTJ. Most devices made using this approach still rely on switching spin currents produced by spin-orbit effects in isotropic channel materials. This places significant challenges to both fundamental materials, structural symmetries, and practical device arrangements due to the orthogonal configuration of spin-current polarization and magnetization. 

To allow deterministic switching, there needs to be a symmetry breaking mechanism to uniquely associate the in-plane current direction to that of the up/down spin polarization.  This can be introduced on a device level through magnetic fields from additional ferromagnetic layers~\cite{8281017,Garello2019,Krizakova2020,Zhao2020} or exchange coupling~\cite{Lau2016,VanDenBrink2016}. Additionally, the free layer switching mechanism changes for this configuration which brings both advantages and disadvantages. Unlike antidamping switching, when the free layer magnetization and spin polarization are not collinear there is no incubation time where the magnetization first slowly begins precessing, allowing for the potential of fast picosecond timescale switching~\cite{Yang2017,Kaushalya2020}. However, this comes at the cost of write current, with current densities for this mechanism nearly two orders of magnitude larger than antidamping switching. Ideally, the write channel should produce a spin current with an out-of-plane spin polarization to decrease the write current. The search for such materials is ongoing with a few materials beginning to show small spin currents of this nature~\cite{2016160}. In addition, trade offs between the requirements for data retentation (i.e. $E_b$), the symmetry-breaking field strength, the write current amplitude, the write speed and the write-error-rate have yet to be established experimentally.

There have also been attempts to introduce some of the advantages from SOT sourced spin currents into the 1T design~\cite{Sato2018,Grimaldi2020,Zhang2021,Wang2018,Jue2018}. In this configuration, free layer switching is driven by spin currents from both a reference layer and SOT channel, indicated in Table~\ref{Table:MM} under the heading STT+SOT, 1T/pMTJ.  This approach has similar characteristics to the other two terminal MTJ devices, such as scalability and the consequences of combining the read/write paths. With the addition of SOT currents, there has been experimental observation of a reduction in switching currents in 2-terminal~\cite{Sato2018} and 3-terminal \cite{Krizakova2020} devices. Further, since the SOT spin current is polarized orthogonal to the free layer magnetization, it may help decrease the incubation time for the free layer. Similar approaches have been considered in MTJ where two reference layers are used with one in-plane and the other perpendicularly magnetized~\cite{Liu2010}. The case of a perpendicularly magnetized free layer with two references layers, one out-of-plane magnetized, the other magnetized in plane has also been reported~\cite{Wolf2020}. However, by using a SOT source the series resistance of a second tunnel barrier can be avoided. 

In recent studies, it has been predicted~\cite{Taniguchi2015b,Belashchenko2019,ochoa2021self} and shown experimentally~\cite{Baek2018,Gibbons2018,iihama2018spin,Seki2019,Safranski2018,Safranski2020} that ferromagnetic materials can be sources of spin current produced by spin-orbit interactions. Unlike the spin-Hall effect in isotropic heavy-metal channel materials (such as Pt) where the polarization is set by geometry, in ferromagnetic materials it is further affected by the direction of the magnetization. In particular, the planar Hall effect has been demonstrated to produce a spin current from laterally flowing charge currents~\cite{Taniguchi2015b,Belashchenko2019,ochoa2021self,Safranski2018,Safranski2020}. More importantly is that the polarization of this current can be partially out of the sample plane, potentially allowing the exertion of antidamping torques on a perpendicular magnetized layer. A 2T design with a PHE injector (SOT-PHE-2T/pMTJ) would have similar characteristics to the injector SOT-2T/pMTJ. However, since there is the potential for an antidamping switching mechanism, the switching current would be expected to be much lower for the SOT-PHE-2T/pMTJ assuming similar charge-to-spin current conversion efficiency. Importantly, in this case, an additional symmetry breaking magnetic field would not be needed. Realization of such a device does bring some additional challenges. Namely, there needs to be materials found with large charge-to-spin conversion and the ability to have magnetization canted at about 45 degrees to the layer plane where a large spin current is expected. Pinning the magnetization at this angle is a materials challenge that would need to be overcome as well. Additionally, it has been shown that excitations in the PHE injector are possible. It is unclear the effect these may have on overall device reliability. 

\vspace{0.5 cm}
\section{Metric and measurement of SOT-generated spin currents}

Measuring SOT-generated spin currents is important for materials understanding and optimization, and is also fundamental for assessing the charge-to-spin current conversion efficiency. 

The most direct metric for spin-current effectiveness is its ability to switch a nanomagnet of interest for MRAM applications. This means a thermally stable nanomagnet ($E_b \gtrsim 60 k_B T$), at realistic operating conditions at the desired switching speed, with low error-rate~\cite{Liu2012,2017139,2017142,2018092,Grimaldi2020,Krizakova2020}. Such characterization needs to be done with channel width and MTJ diameter well below $50$~nm, necessitating demanding lithography and sample patterning, which slows materials and device exploration pace and turn-around time. Various alternative measurements have been employed in SOT-related studies that do not require elaborate device structure fabrication. However, a quantitative connection from these measurements to the device performance has not been unambiguously established~\cite{Zhu2021}.

Besides switching experiments, two common ways of measuring the strength of an antidamping spin-current are: (1) by spin torque ferromagnetic resonance (ST-FMR)~\cite{2012065}; and (2) methods based on  the inverse spin-Hall effect~\cite{2012116}. In both cases, multiple steps of spin-charge conversions are involved. Also, the spin current being measured usually crosses one or more materials interfaces. 

A third frequently reported method for measuring SOT-related spin current is the use of the anomalous and planar Hall effects to characterize magnetization response to spin currents. The technique relies on measuring a nonlinear Hall response, a so-called second harmonic Hall signal~\cite{2021175,2013110,2014008,2021174,2015160}. This is an accessible and convenient measurement. However, the second harmonic signal contains many components, including that of an interface Rashba-field-like term, an often uncontrolled Joule heating, as well as possibly from details of interface magnetic moment's anisotropy potential~\cite{Safranski2019,2021172}. These together add complexity to the quantitative determination of the charge-to-spin conversion efficiency.

For materials characterization and for fundamental physics, one often describes the charge-to-spin conversion ratio by a simple materials parameter---called the spin-Hall angle $\theta_\mathrm{SH} $~\cite{Hirsh1999,Zhang2000,Kimura2007,2013058,2021171,2012065}. While conceptually well defined, the spin-Hall angle $\theta_\mathrm{SH}$ is not always directly relatable to observables. In most practical materials and device physics experiments, the SOT generated spin current needs to traverse materials interfaces, where the transport and spin-flip scattering can add complexity, both to mechanisms and to poorly controlled materials related parameters even when the mechanisms are known. 
 
Measurement of the free layer moment's dynamics could also become uncertain, one aspect being the nanomagnet's interface magnetic moment may experience different anisotropy and exchange-coupling environment than those in free layer interior~\cite{2018006,Safranski2019}. Since the spin currents we consider produce a spin-torque within a few atomic spacings of an interface, the spin current's action on a free layer may further be affected by interface-related inhomogeneous spin-excitation processes. The materials physics origins of many such measurement inconsistencies are being actively investigated~\cite{2021172,2020011,Xu2020,2021046}.

\section{\label{Sec:Outlook}Outlook and Perspective}
The exploration of spin-current generation and propagation has seen great scientific progress over the past decade. Commercial applications based on STT-MTJ are appearing. SOT-based devices are being actively explored for faster, higher performance memory technologies, possibly for direct integration with processor logic circuit, known as ``embedded'' memories~\cite{2021170}, such as for last few levels of cache memory in a CPU chip, replacing the high-performance but volatile and expensive static random access memory (SRAM). To capture these advantages, one must have economical integration of spintronic devices into existing CMOS technology nodes. An important consideration is the matching of charge transport characteristics between the spintronic devices and that of CMOS circuits at targeted technology nodes. Two key attributes for a spintronic device to optimize towards are (1) the amount of charge current for write should be sufficiently low to match what a transistor can source, and (2) the MTJ for read-out needs to be impedance matched with the read and write circuit. 

\subsection{CMOS transistor characteristics}
\begin{figure}[tbp!]
\includegraphics[width=3.2in]{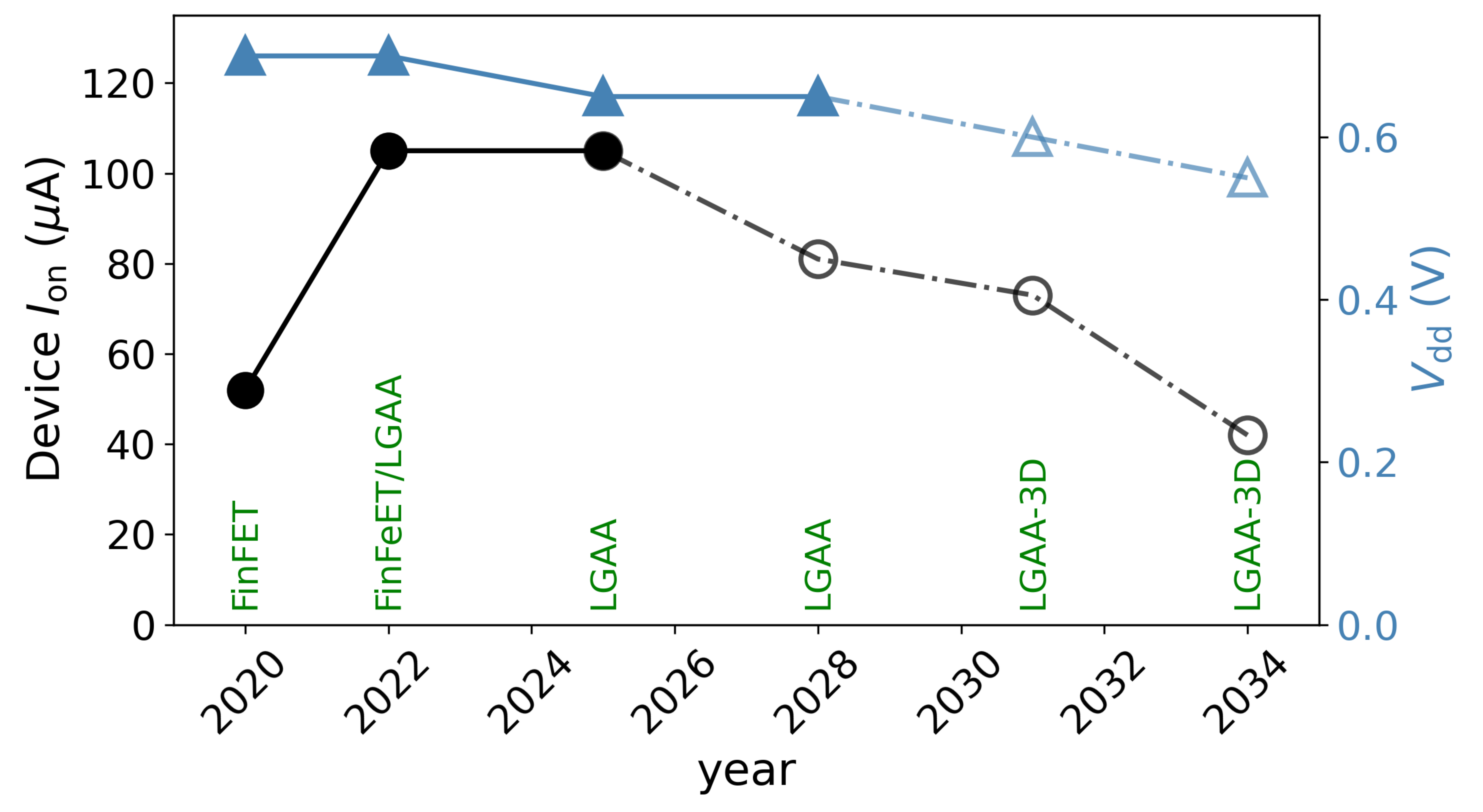}
\caption{A projection of CMOS-transistor on-state current (left y-axis) and supply voltage $V_\mathrm{dd}$ (right y-axis) over the next decade, according to IRDS 2020's ``More Moore'' report~\cite{IRDS2020}. Labels in green describe the device structure technology scenarios. Data in open symbols are more speculative.}
\label{Fig:CMOS}
\end{figure}

Figure~\ref{Fig:CMOS} shows the IRDS (2020) traditional CMOS transistor on-state source-drain current over the next decade. While transistor technology advances, such as with vertical field effect transistors that may increase the write current available, these traditional CMOS-transistor values still provide a good benchmark for STT-MRAM scaling. To satisfy the on-state current and voltage for such ``write'' operation, one needs to reduce both the switching current, and switching current-path’s resistance. 
This demand motivates much materials and device physics exploration for increasing charge-to-spin current conversion efficiency, beyond that possible with spin-filtering in a MTJ.

\subsection{SOT devices: beyond spin-filtering}
 SOT-based devices can produce better efficiency, both theoretically and per experimental demonstration, especially when using strong spin-orbit-interaction compounds such as the so-called topological insulators, where $I_\mathrm{s}/I_\mathrm{cg}  > 1$ has been demonstrated~\cite{2014136,2014137}. The challenges for these materials are (1) their growth and processing compatibility with CMOS technology, and (2) a limited tolerance in applications to the SOT channel resistance, due to the above-mentioned matching requirement to CMOS environment. A scaled up practical integration of such SOT-devices based on topological insulators materials has yet to be demonstrated.

Another important factor for SOT-device applications is its spin-current's polarization direction. As discussed above, pMTJs are often necessary for high area-density memory technology. For such devices, an SOT-based spin-current augmentation require its polarization to have a large component perpendicular to the channel film plane. This is particularly important for the STT+SOT 1T/pMTJ type in Table~\ref{Table:MM}, since the density requirement remains premium. While for 2T/pMTJ applications, often there is a relaxation to the density requirement in trade for higher performance. Therefore, for 2T/MTJ structures, it is reasonable to consider possible applications with in-plane magnetized MTJs~\cite{2016118,2017139}, if it is able to demonstrate superior switching speed at acceptable switching current, while also providing relatively simple integration routes. 

Since the MTJ functions as a read-out element, its resistance also needs to be matched to the read-out circuit's input impedance, and cannot be arbitrarily reduced. In this context, a 3-terminal SOT-based device allows for separate optimization of the read- and write-circuit's impedance, which is an advantage, to be weighed against increased memory cell unit area, which translates to memory density, and hence cost of the technology.  

\subsection{Materials compatibility}
Materials compatibility is an important consideration for integrated technology offering such as high performance MRAM. Initial results show that SOT-based devices can be integrated with CMOS technology~\cite{Garello2019VLSI}, but more work is needed. As the same time, STT-MTJ based devices appear to continue to provide advanced functionality in a cost-effective fashion for future technology nodes.

In the case of SOT-devices, a key materials factor is the understanding and control of properties of the spin-current-crossing interfaces. The interface needs to be robust and reproducible for manufacturability, and it needs to be compatible to the established CMOS manufacturing thermal and processing conditions. This typically requires enduring a thermal budget of 300-400 C for up to an hour. 

For producing perpendicularly polarized spin-current through materials innovation, one needs to include the discussion of in-plane symmetry control for such SOT materials at an early stage. For example, non-magnetic materials with a lack of crystal inversion symmetry can produce perpendicularly polarized spins~\cite{2016160,2021160}. However, methods have to be developed to control their crystal orientation on an amorphous or polycrystalline base-layer --- the type of substrate surfaces typically encountered in back-end CMOS integration environment. An advantage of using ferromagnetic materials themselves to set a spin polarization direction is the magnetization sets the spin-polarization directions. The challenge---as discussed above (Sec.~\ref{Sec:SCS})---is to set and fix a magnetization direction that is neither in-plane nor perpendicular to the plane, which is required for these spin currents that flow perpendicular to the plane of the layers.

SOT-induced antidamping spin-current encompasses a rich set of phenomena involving both bulk materials properties and its interface-specific structures and behaviors, producing a wealth of knowledge to the materials and their combination's behavior. For applications to memory technology, the focus is on increasing charge-to-spin current conversion ratio, and on ensuring at the same time the feasibility for materials integration with existing and future CMOS technology. While it provides a promising path forward to high-density and high-performance MRAM and other magnetic devices, there is no shortage of challenging engineering problems to solve going forward.

\subsection{Open Scientific Questions}
There are many basic open scientific questions related to the spin-orbit torque mechanisms of charge-to-spin conversion. Spin-orbit coupling comes into determining a material's electronic structure and is important in electron scattering processes. In the first case, the conversion process is considered intrinsic to the material, derived from a material's band structure. Whereas in the second case, it is considered extrinsic, associated with impurity or other electron scattering mechanisms. Fundamental questions thus relate to identifying the important conversion processes, and, ideally, a dominant contribution. Further, spin-orbit coupling at an interface is different from that in the bulk of the materials. Most notably, the spin polarization of interface-generated spin currents is not necessarily in the plane of the interface or bound by crystal symmetry constraints~\cite{Amin2018,Amin2020}. 

Spin orbit torques associated with magnetically ordered materials are also being be explored now in greater depth. The spin polarization direction within a ferromagnetic material is collinear with the magnetization direction. However, again interface and bulk scattering will both influence the spin polarization. For example, spin polarization generated by the anomalous or planar Hall effects need to cross a ferromagnetic/nonmagnetic interface at which spin filtering and spin-orbit precession can occur~\cite{Amin2018,Amin2020}. 
We also note that there are interesting ways that spin direction and the direction of charge flow can be interchanged, effects again associated with spin-orbit coupling known as spin swapping~\cite{Lifshits2009} or spin rotation~\cite{Aljuaid2018,Humphries2017,Luo2019}. 
 There are additionally ideas for generating such spin polarizations using antiferromagnets~\cite{Gibbons2021}, chiral antiferromagnets~\cite{Felser2017}, and non-collinear antiferromagnets, e.g. Mn$_3$Ir~\cite{WZhang2016}. Further, first-principles calculations have shown that spin torques with a planar Hall effect symmetry can be associated with interfaces, e.g. Co$|$Pt~\cite{Belashchenko2019,Belashchenko2020a,Belashchenko2020b}. It is clear that experiment and theory are advancing field rapidly at the time of this writing and that there will continue to be interesting predictions and experimental results.

\section{\label{Sec:Summary}Summary}

In summary, spin-current generation has become an important research topic in recent years, with significant progress being made in increasing the efficiency of charge-to-spin conversion processes. Further, it is now appreciated that controlling the spin-polarization direction is as important as increasing the efficiency. This perspective article has identified several critical research needs. First, the development of channel materials with larger charge-to-spin conversion efficiency, based ideally, on a microscopic understanding of the conversion mechanisms that enables material optimization. For example, conversion that relies on intrinsic mechanisms may require cleaner, more ordered channel materials and interfaces. Those based on extrinsic mechanisms may require controlled material doping to introduce particular types of electron scattering sites. Second, that the spin accumulations created have a spin polarization component perpendicular to the channel interface to enable devices based on sub-20 nm diameter perpendicularly magnetized MTJ nanopillars without the need for a symmetry breaking field acting on the free layer magnetization. Third, to have an important impact on semiconductor memory technology, the resulting devices and structures must be CMOS compatible, in terms of materials, materials processing and CMOS-transistor electrical characteristics (e.g. shown in Fig.~\ref{Fig:CMOS}). Finally, metrology methods must be developed that can benchmark channel materials before nanopillar device fabrication and testing, to accelerate materials development. With the worldwide research focus on this and closely related topics, it appears clear that there will continue to be important advances in this field that can drive technological advances in MRAM and beyond.

\section*{Acknowledgments}
This work was supported in part by NSF DMR-2105114. We thank Egecan Cogulu for creating Fig.~\ref{Fig:Orthogonal} and the illustrations in Table~\ref{Table:MM}.
\end{document}